\documentstyle[12pt,epsfig]{article}
\newlength{\dinwidth}
\newlength{\dinmargin}
\newcommand{\ba}{\begin{array}}
\newcommand{\ea}{\end{array}}
\newcommand{\beq}{\begin{equation}}
\newcommand{\eeq}{\end{equation}}
\newcommand{\bea}{\begin{eqnarray}}
\newcommand{\eea}{\end{eqnarray}}

\def\S{{\bf S}}

\def\bce{\begin{center}}
\def\ece{\end{center}}

\def\nonu{\nonumber}

\def\pa{\partial}
\def\al{\alpha}
\def\be{\beta}
\def\ga{\gamma}

\def\de{\delta}
\def\De{\Delta}

\def\la{\lambda}
\def\La{\Lambda}

\def\si{\sigma}

\def\S{{\bf S}}

\begin{document}
\thispagestyle{empty}
\addtocounter{page}{-1}
\begin{flushright}
SNUST 99-008\\
{\tt hep-th/9911199}\\
\end{flushright}
\vspace*{1.3cm}
\centerline{\Large \bf More CFTs and RG Flows}
\vskip0.3cm
\centerline{\Large \bf from}
\vskip0.3cm
\centerline{\Large \bf  Deforming M2/M5-Brane 
Horizon~\footnote{ Work supported 
in part by the BK-21 Initiative Physics Program, 
KRF International Collaboration Grant and the KOSEF Interdisciplinary 
Research Grant 98-07-02-07-01-5. }}
\vspace*{1.0cm} 
\centerline{\bf Changhyun Ahn${}^a$ {\rm and} Soo-Jong Rey${}^{b}$}
\vspace*{0.8cm}
\centerline{\it ${}^a$ Department of Physics, 
Kyungpook National University, Taegu 702-701 Korea}
\vskip0.3cm
\centerline{\it ${}^b$ School of Physics, Seoul National University,
Seoul 151-742 Korea}
\vspace*{0.8cm}
\centerline{\tt ahn@kyungpook.ac.kr, \hskip0.5cm sjrey@gravity.snu.ac.kr}
\vskip1.5cm
\centerline{\bf abstract}
\vspace*{0.5cm}
Near-horizon geometry of coincident M2-branes at a conical singularity is 
related to M-theory on $AdS_4$ times an appropriate seven-dimensional 
manifold $X_7$. 
For $X_7=N^{0, 1, 0}$, squashing deformation is known to lead to 
spontaneous (super)symmetry breaking from ${\cal N}=(3, 0)$ to ${\cal N}=(0, 1)$
in gauged $AdS_4$ supergravity. Via AdS/CFT correspondence, it is interpreted 
as renormalization group flow of strongly coupled three-dimensional field
theory with $SU(3) \times SU(2)$ global symmetry. The flow interpolates
between ${\cal N}=(0, 1)$ fixed point in the UV to ${\cal N}=(3, 0)$ fixed 
point in the IR.
Evidences for the interpretation are found both from critical points of the 
supergravity scalar potential and from conformal dimension of relevant chiral
primary operators at each fixed point. We also analyze cases with 
$X_7=SO(5)/SO(3)_{\rm max}, V_{5,2}(R), M^{1,1,1}, Q^{1,1,1}$ and find that
there is no nontrivial renormalization group flows. We extend the analysis 
to Englert type vacua of M-theory. By analyzing de Wit-Nicolai 
potential, we find that deformation of $\S^7$ gives rise to renormalization 
group flow from ${\cal N}=8$, $SO(8)$ invariant UV fixed point to 
${\cal N}=1$, $G_2$ invariant IR fixed point. 
For $AdS_7$ supergravity relevant for near-horizon geometry of coincident
M5-branes, we also point out a nontrivial renormalization group flow from 
${\cal N}=1$ superconformal UV fixed point to non-supersymmetric IR fixed 
point.

\baselineskip=18pt
\newpage

\section{Introduction}
\setcounter{equation}{0}
Few examples are known for three-dimensional interacting conformal field 
theories or for renormalization group flows among themselves, mainly due to 
strong coupling dynamics in the infrared limit. For example, ${\cal N}= 1$ 
superconformal field theories, which arise in various situations involving
D-brane dynamics, do not have any continuous R-symmetries, holomorphy 
constraints or any known non-renormalization theorems. 
In the previous paper \cite{ar}, as an alternative route, we have proposed to  
classify three-dimensional (super)conformal field theories by utilizing 
the AdS/CFT correspondence \cite{maldacena,witten,gkp} and earlier, exhaustive 
study of the Kaluza-Klein supergravity \cite{duff}.

The simplest spontaneous compactification of the eleven-dimensional 
supergravity \cite{cremmer} is the Freund-Rubin \cite{fr} compactification
\footnote{corresponding to the near-horizon geometry of coincident, infinitely 
planar M2-branes} to a product 
of $AdS_4$ spacetime and an arbitrary compact seven-dimensional Einstein 
manifold $X_7$ of positive scalar curvature. Continuous deformations among 
$X_7$'s is of interest 
\footnote{We restrict our foregoing discussions only to deformations among
$X_7$'s of {\sl same} topology. While examples of deformation between $X_7$'s
of {\sl distinct} topology would be extremely interesting, we are not aware
of any explicit example.}, as they would be interpreted, in the strongly 
interacting $d=3$ quantum field theory dual to the $AdS_4$ supergravity, to 
renormalization group flows among interacting (super)conformal fixed points.
The best known example is provided by round and squashed $\S^7$. The standard 
Einstein metric of the round $\S^7$ yields a vacuum with $SO(8)$ gauge 
symmetry and ${\cal N}=8$ supersymmetry. The $\S^7$ also admits the second, 
squashed Einstein metric \cite{jensen}, yielding a vacuum with $SO(5) \times 
SO(3)$ gauge symmetry and ${\cal N}=1$ or ${\cal N}=0$ supersymmetry, depending on the orientation of the $\S^7$ \cite{awada}.

In \cite{ar}, we have shown that the well-known spontaneous (super)symmetry 
breaking deformation from round- to squashed-$\S^7$ is mapped to a 
renormalization group flow from ${\cal N}=0$ or 1, $SO(5) \times SO(3)$ 
invariant fixed point in the UV to ${\cal N}=8$, $SO(8)$ invariant fixed 
point in the IR. In particular, in \cite{ar}, 
we have shown that (1) the squashing 
deformation corresponds to an irrelevant operator at the ${\cal N}=8$ 
superconformal fixed point and a relevant operator at the ${\cal N}=1$ or 0 
(super)conformal fixed point, respectively, and (2) the renormalization group 
flow is described geometrically, as in $AdS_5$ supergravity \cite{domainwall}, 
by a `static' domain wall of the sort studied earlier in \cite{cgr}, 
which interpolates the two asymptotically $AdS_4$ spacetimes with $X_7$ round 
and squashed $\S^7$'s.

In this paper, we will be studying further known examples of Kaluza-Klein 
supergravity vacua and reinterpret them in terms of three-dimensional
(super)conformal field theories and associated renormalization group flows. 
First, we will be exploring various Freund-Rubin type spontaneous
compactifications on $AdS_4 \times X_7$. For M2-branes on an eight-dimensional
manifold, the near-horizon geometry $X_7$ is expected to change as the 
branes are placed at or away a conical singularity of the manifold \cite{qmw, 
morrisonplesser}. More specifically, we will consider $X_7$ being
1) 3-Sasaki, 2) Sasaki-Einstein or 3) weak $G_2$ holonomy manifolds,
describing near-horizon geometry of M2-branes at relevant conical
singularities.
Each of them preserves $(3,0), (2,0)$ and $(1,0)$ supersymmetries,
respectively, where $(N_L, N_R)$ denotes chirality of supercharges on $AdS_4$.

Explicit examples of the above $X_7$'s are

1) $N^{0,1,0}_{\rm I}$ space with ${\cal N}=(3, 0)$

2) $M^{1,1,1}, Q^{1,1,1}, V_{5, 2}(\rm R)$ spaces with ${\cal N}=(2,0)$

3-L) Squashed $\S^7, SO(5)/SO(3)_{\mbox{max}}$ and other $N^{p,q,r}_
{\rm I}$ space with ${\cal N}=(1, 0)$

3-R) all $N^{p,q,r}_{\rm II}$ space with ${\cal N} = (0, 1)$ .

\noindent In section 2, by analyzing relevant scalar potentials developed by 
Yasuda \cite{yasuda,yasuda1} for each case, we will be finding further
examples of supergravity dual to a three-dimensional quantum field theory 
exhibiting a nontrivial renormalization group flow.

The first example is provided by $X_7 =  N^{0, 1, 0}_{\rm I, II}$.
The manifold $N^{p,q,r}_{\rm I}$ has been studied originally by Castellani 
and Romans \cite{cr}, 
identified as a coset manifold of the form $[SU(3) \times U(1)]
/[U(1)\times U(1)]$.  Embedding of $[U(1) \times U(1)]$ in $[SU(3)\times U(1)]$ 
is specified by  a choice of the integers $(p,q,r)$.
Each choice of $p,q,r$ leads to an Einstein manifold, yielding ${\cal N}=(3,0)$ 
supersymmetry and $SU(3)\times SU(2)$ gauge symmetry for $N^{0,1,0}$, or 
${\cal N}=(1,0)$ supersymmetry and $SU(3)\times U(1)$ gauge symmetry otherwise. 
Later, Page and Pope \cite{pp} have completed the coset manifold construction 
by showing existence of another series of Einstein manifold, 
$N^{p,q,r}_{\rm II}$, except for those corresponding to $q=0$. 
The $N^{p,q,r}_{\rm II}$'s, which can be obtained from geometric squashing 
of the $N^{p,q,r}_{\rm I}$'s, retain the same gauge group as the 
latter but instead preserve ${\cal N} = (0,1)$ supersymmetry.

As in the case of $X_7 = \S^7$, the existence of two Einstein metrics on
$N^{p,q,r}$'s  offers an interpretation in terms of spontaneous (super)symmetry 
breaking: scalar field corresponding to the squashing deformation acquires a 
nonzero vacuum expectation value, leading to (super)-Higgs mechanism. 
The symmetry breaking pattern depends on the choice of the $X_7$ orientation.
With one orientation, for $N^{0,1,0}$'s, the squashing interpolates between a 
${\cal N}=3$ supersymmetric vacuum and another with ${\cal N}=0$ supersymmetry. 
With opposite orientation, it interpolates between a non-supersymmetric
vacuum and a supersymmetry restored one with ${\cal N}=1$ supersymmetry.
In both cases, however, the gauge group is locally $SU(3) \times SU(2)$
and remains unbroken.

Via AdS/CFT correspondence, this implies that the three-dimensional quantum 
field theory dual to the
$AdS_4 \times N^{0,1,0}$ supergravity exhibits two types of nontrivial
renormalization group flow: one between ${\cal N}=3$ superconformal UV fixed 
point and non-supersymmetric IR fixed point, and another between 
nonsupersymmetric UV fixed point and ${\cal N}=1$ superconformal IR fixed
point. For both, along the renormalization group flow trajectory, the global
symmetry $SU(3) \times SU(2)$ is always maintained. Analogously, for all other 
$N^{p,q,r}$'s except $q = 0$, the dual quantum field theory exhibits two 
renormalization group flows between ${\cal N}=1$ superconformal fixed point
and nonsupersymmetric one. We present these analysis in Section 2.1.

In Section 2.2, we will be studying other examples of $X_7$'s.    
We will find that $N^{p,q,r}_{\rm I, II}$'s give rise to a nontrivial
, $SU(3) \times U(1)$ invariant renormalization group flow between 
${\cal N}=1$ superconformal fixed point and nonsupersymmetric conformal fixed 
point. On the other hand, for $X_7 = M^{1,1,1}, Q^{1,1,1}, V_{5,2}({\rm R}), 
SO(5)/SO(3)_{\rm max}$, it turns out the squashing deformation does not lead 
to any nontrivial renormalization group flow \footnote{Other aspects of 
three-dimensional superconformal field theories from Sasaki seven-manifold
have been studied in \cite{fabbri}.}. 

We next consider M-theory compactification vacua of Englert type. By 
generalizing compactification vacuum ansatz to the nonlinear level,
solutions of the eleven-dimensional supergravity were obtained 
directly from the scalar and pseudo-scalar expectation values at
various critical points of the ${\cal N}=8$ supergravity potential \cite{dnw}. 
This way, it was possible to reproduce all known Kaluza-Klein solutions of 
the eleven-dimensional supergravity: round $\S^7$ \cite{dp}, 
$SO(7)^-$ invariant, {\sl parallelized} $\S^7$ \cite{englert},
$SO(7)^{+}$-invariant vacuum \cite{dn}, $SU(4)^{-}$-invariant vacuum 
\cite{pw}, and a new one with $G_2$ invariance. 
Among them, round $\S^7$- and $G_2$-invariant vacua are
stable, while $SO(7)^{\pm}$-invariant ones are known to be unstable 
\cite{dn1}. In all these vacua, generically, either the metric is 
endowed with a nontrivial {\sl warp factor} or the four-form {\sl magnetic} 
flux $G_{abcd}$ is nonvanishing, corresponding to turning on vacuum 
expectation values for scalar or pseudoscalar field, respectively.  
Novelty of vacua with a nontrivial warp factor is that they corresponds to 
{\sl inhomogeneous} deformations of $\S^7$. 
In section 3, we will analyze the above vacua by investigating 
de Wit-Nicolai potential. We will be identifying a deformation which gives 
rise to a renormalization group flow associated with the symmetry breaking 
$SO(8) \rightarrow G_2$ (both of which are stable vacua) 
and find that the deformation operator is relevant at the $SO(8)$ 
fixed point but becomes irrelevant at the $G_2$ fixed point.

In Section 4, we study near-horizon geometry of coincident M5-branes, 
which are described by $AdS_7$ gauged supergravity. 
The geometrical interpretation of seven dimensional solutions from 
the eleven dimensional viewpoint has been obtained recently \cite{lp}.
By taking a nonlinear Kaluza-Klein ansatz to the eleven-dimensional 
supergravity, they were able to obtain a consistent $\S^4$ reduction
to seven-dimensional, ${\cal N}=1$ gauged supergravity. 
Field content of the latter includes a scalar field parametrizing 
{\sl inhomogeneous} deformations of the $\S^4$, Yang-Mills gauge fields,
and a topologically massive three-form potential. Turning on the scalar 
field induces again spontanous (super)symmetry breaking, and we interpret
it as being dual to a renormalization group flow from ${\cal N}=1$ 
supersymmetric UV fixed point to a nonsupersymmetric IR fixed point of a 
putative $d=6, {\cal N} = (1,0)$ quantum theory.
The existence of stable, nonsupersymmetric $d=6$ conformally invariant
quantum theory should be of considerable interest \cite{berkoozrey}, as the latter includes 
{\sl noncritical bosonic strings} as part of the spectrum.

\section{ 3d CFTs from Freund-Rubin Compactifications}
\setcounter{equation}{0}

Spontaneous compactification of M-theory to $AdS_4 \times X_7$ is 
obtained from near-horizon geometry of $N$ coindicent M2-branes. 
The configuration is equivalent to Freund-Rubin compactification of
the eleven-dimensional supergravity: through $X_7$, the M2-branes thread 
nonvanishing flux of four-form field strength
\bea
G_{\alpha \beta \gamma \delta}
= \frac{1}{\sqrt{-g_4}} Q e^{-7 s } \epsilon_{\alpha \beta \gamma \delta},
\hskip1.5cm Q \equiv 96 \pi^2 N \ell_{\rm pl}^6.
\label{fourform}
\eea
The parameter $Q$ refers to so-called `Page' charge \cite{page},
 $Q \equiv \pi^{-4} \int_{X^7} ({}^*G + C \wedge G)$.
Here, the $d=11$ coordinates with indices $A, B, \cdots$ are decomposed into 
$AdS_4$ coordinates $x$ with indices $\al, \be, \cdots $ and 
$X_7$ coordinates $y$ with indices $a, b, \cdots$. We will also adopt 
the eleven-dimensional metric convention as $(-, +, \cdots, +)$. 
A Weyl rescaling of $AdS_4$ and $X_7$, appropriate for M-theory description,  
is given by $g_{\al \be} \rightarrow e^{7s} g_{\al \be}$
and $g_{ab} \rightarrow e^{-2s} g_{ab}$.

In this section, we will be studying critical points of the scalar potential
for various choices of $X_7$ and, in the corresponding three-dimensional
conformal field theories, identifying an operator that gives rise to a 
renormalization group flow among the critical points. 

\subsection{$N^{0,1,0}$ space}
We will begin with the case $X_7 = N^{p,q,r}$ space \cite{cr}, which is a 
homogeneous space $[SU(3) \times U(1)]/[U(1) \times U(1)]$.
Consider a subset of all continuous deformation of $N^{p,q,r}$, in which
the vielbein $B^{\al}$ is given by
\bea
B^{\al}=(\frac{1}{\al} \Omega^a, \frac{1}{\be}( p \Omega^8+q \Omega^3 +
r \Omega^{3'}), \frac{1}{\ga} \Omega^{A}, \frac{1}{\delta} \Omega^{A'}) .
\nonu
\eea
Here, $\Omega^{a}( a=1, 2), \Omega^3, \Omega^{A} (A=4, 5), \Omega^{A'}(A'=6,
7), \Omega^8$ and $\Omega^{3'}$ are the left invariant one-forms
corresponding to the coset generators $\la^a, \la^3, \la^A, \la^{A'}, \la^8$
of $SU(3)$ and the generator of $U(1)$ factor respectively. The integer
$p, q, r$ characterize the embedding of $U(1) \times U(1)$ in $SU(3) \times 
U(1)$.
It is convenient to define  
\bea
\al=e^{s+u/2-v}, \quad \be=e^{s-3u}, \quad \ga=e^{s+(u+v+w)/2}, \quad 
\de=e^{s+(u+v-w)/2}.
\label{suvw}
\eea
The deformation considered above can be summarized compactly in terms of
four-dimensional effective Lagrangian \cite{yasuda}
\bea
{\cal L}_4=\frac{1}{2} \sqrt{-g_4} \left( -R_4-\frac{63}{2} \left(
\pa_{\mu} s \right)^2-\frac{21}{2} \left(\pa_{\mu} u \right)^2-
3 \left( \pa_{\mu} v \right)^2 -\left( \pa_{\mu} w \right)^2 -
V(s,u,v,w) \right) ,
\nonu
\eea
where 
\bea
V(s,u,v,w) & = & e^{9s} \left( -\frac{3}{2} e^u \left( e^{-2v}+
e^{v+w} + e^{v-w} \right) +\frac{1}{4} e^u \left( e^{-2v+2w}+
e^{-2v-2w} +e^{4v} \right) \right. \nonu \\
& & \left. + \frac{1}{4} e^{8u-4v}   + \frac{(1+x)^2}{16}
e^{8u+2v+2w} +\frac{(1-x)^2}{16} e^{8u+2v-2w} \right)+Q^2 e^{21s}
\nonu
\eea
and $Q$ is the aforementioned Page charge. 
Critical points of the deformation are determined
by the stationarity condition $\frac{\pa V}{\pa u}=\frac{\pa V}{\pa v}=
\frac{\pa V}{\pa w}=0$ \footnote{Actually this is equivalent to the 
requirement \cite{cr}
that the internal 7 manifold has to be an Einstein space. The rescalings 
are fixed by this condition \cite{cr}.}. 
It turns out the solution \cite{cr,castellani}  is specified by one parameter 
$c$ \footnote{The normalization convention we adopt here is as follows:
$R^{\al}_{\be}= -24 e^2 \de^{\al}_{\be}, R^a_b= 12 e^2 
\de^a_b$ and $G_{\al \be \ga \de}= e \epsilon_{\al \be \ga \de}$. } :
\bea
& & \al^2=\frac{64}{(c+2)^2} (\frac{5}{4} c^2+3c+2) e^2, \quad;
\be=\pm \frac{16(\frac{5}{4} c^2+3c+2)}{(c+2)(3c+2)} e, \nonu \\
& & \ga^2=\frac{64}{(c+2)^2} (c+\frac{d}{2}+\frac{3}{2} ) e^2, \qquad 
\de^2=\frac{64}{(c+2)^2} (c-\frac{d}{2}+\frac{3}{2} ) e^2 ,
\label{cons}
\eea
with 
$-1 \leq c \leq 1$ and $d=\pm \sqrt{1-c^2}$ and $c$ is related to
\bea
x \equiv \frac{3p}{q}=-\frac{(5c+6)d}{3c+2}.
\label{x}
\eea
It is known \cite{cr}
that for a particular choice of $p, q, r$ the isometry of 
$N^{p, q, r}$ is $SU(3) \times SU(2)$ in which ${\cal N}=3$ supersymmetry
survives. We call it type I solution.
For all the other values of $p, q, r$, supersymmetry is broken to
${\cal N}=1$ and the isometry group is $SU(3) \times U(1)$.
Moreover, as we mentioned in the introduction, 
Page and Pope worked out a different construction of metrics on the
$N^{0,1,0}$ space and showed that there exists an another Einstein metric
, squashed from the first. The isometry is $SU(3) \times SU(2)$
and the supersymmetry is ${\cal N}=(0, 1)$. We call this type II solution.
Strictly speaking, the analysis of Killing spinors in \cite{pp} shows that
a possible different root of squashing parameter gives different 
supersymmetry. In other words, Type I solutions are known to preserve
$(3, 0)$. Similarly, the case of 
Type II solutions have $(0, 1)$ supersymmetry. Thus for
the squashing with left-handed orientation, the renormalization group flow
interpolates between the boundary conformal field theories with ${\cal N}=3$ 
and ${\cal N}=0$ supersymmetry while for the squashing with right-handed 
orientation, the flow interpolates between conformal field theories with
${\cal N}=0$ and ${\cal N}=1$.
From now on we will restrict foregoing discussions to $(p,q,r) = (0,1,0)$ 
case.

\begin{figure}[htb]
\label{fig1}
\vspace{0.5cm}
\epsfysize=8cm
\epsfxsize=8cm
\centerline{
\epsffile{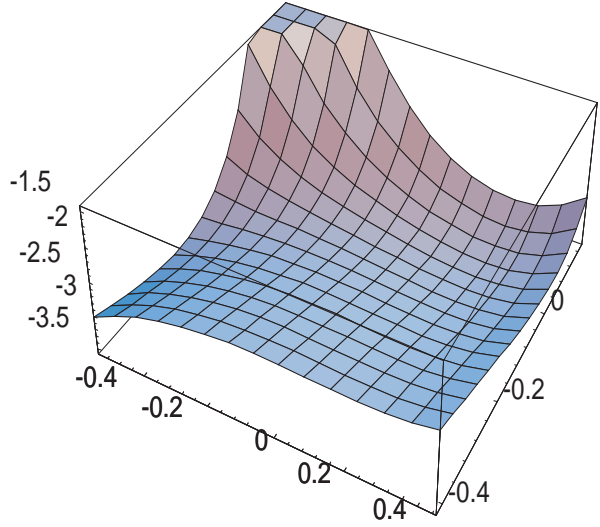} }
\vspace{0.5cm}
\caption{\sl (u,v)-subspace slice of the $N^{0,1,0}$ scalar potential
over $u = [-0.45,+0.15]$, $v = [-0.5, + 0.5]$. The Type I vacuum is the
saddle point in the middle-left corner, and the Type II 
vacuum is the local minimum in the middle-right corner.}
\end{figure}
\vspace{0.5cm}

\vskip0.3cm
$\bullet$ \tt Type I solution for the space $N_{\rm I}^{0,1,0}$
\rm
\vskip0.3cm

Type I solution with ${\cal N} = (3,0)$, as studied by Castellani and 
Roman \cite{cr}, is defined by
\bea
-1 \leq c \leq -\frac{2}{\sqrt{5}}, \;\;\; d=\sqrt{1-c^2}. 
\nonu
\eea
In this case, the Eq. (\ref{cons}) become  
\bea
\al^2=16e^2, \;\; \be= \pm 4e, \;\; \ga^2=32e^2, \;\; \de^2=32e^2
\label{abgd}
\eea
where we put $c=-1$ and $d=0$ by plugging $p=0$ into Eq. (\ref{x}).
Differentiating $V$ with respect to $u,v,w$ with $x=0$, we obtain
\bea
& & V_{uu}|= 21\times 2^{3/7} e^{9s_{\rm I}}, \hskip0.5cm  
V_{vv}|=6 \times 2^{3/7} e^{9s_{\rm I}}, \hskip0.5cm 
V_{ww}| =-2^{3/7} e^{9s_{\rm I}}, \nonu \\
& & V_{uv}|=0, \hskip0.5cm V_{uw}|=0, \hskip0.5cm V_{vw}|=0,
\nonu
\eea
evaluated at $s_{\rm I}, u_{\rm I}, v_{\rm I}$ and $w_{\rm I}$ 
which are the extremum values corresponding to field $s, u, v,$ and $w$, 
respectively,
\bea
e^{12s_{\rm I}}=\frac{9}{2^{18/7}} \frac{1}{Q^2}, \hskip0.5cm 
u_{\rm I}=\frac{2}{21} \ln 2, \hskip0.5cm v_{\rm I}= \frac{1}{3} \ln 2, 
\hskip0.5cm w_{\rm I}=0
\nonu
\eea
that can be obtained explicitly using Eq.(\ref{suvw}) and Eq.({\ref{abgd}).

Conformal dimension of the perturbation operator representing squashing 
is determined by fluctuation spectrum of the scalar fields. After rescaling
$\sqrt{63} s = \overline{s}, \sqrt{21} u =\overline{u}, 
\sqrt{6} v = \overline{v}, 
\sqrt{2} w = \overline{w} $, one
finds that the nonzero 
fluctuation spectrums for $u, v$ and $w$ fields around the
$N^{0,1,0}_{I}$ take 
\bea
M_{\overline{u} \overline{u}}^2(N^{0,1,0}_{\rm I}) 
& = & \left[ \frac{\pa^2 V}{\pa \overline{u}^2}\right]_{
\overline{s_{\rm I}}, \overline{u_{ \rm I}}, \overline{v_{\rm I}},
\overline{w_{\rm I}}}
 = \qquad   + 2^{3/7} e^{9s_{\rm I}}, \nonu \\
  M_{\overline{v} \overline{v}}^2(N^{0,1,0}_{\rm I}) & = & 
\left[ \frac{\pa^2 V}{\pa \overline{v}^2}\right]_{
\overline{s_{\rm I}}, \overline{u_{\rm I}}, \overline{v_{\rm I}},
\overline{w_{\rm I}}}
 =  
\qquad +2^{3/7} e^{9s_{\rm I}},  \nonu \\
  M_{\overline{w} \overline{w}}^2(N^{0,1,0}_{\rm I}) & = & 
\left[ \frac{\pa^2 V}{\pa \overline{w}^2}\right]_{
\overline{s_{\rm I}},
\overline{u_{\rm I}}, \overline{v_{\rm I}},
\overline{w_{\rm I}}}
 = 
\qquad - \frac{1}{2} \times
 2^{3/7} e^{9s_{\rm I}}.   
\nonu
\eea
The cosmological constant $\Lambda_{\rm I}$ is
\bea
\La_{\rm I}= \frac{V_{\rm I}}{2} = - \frac{9 \sqrt{6}}{
16} \left( \frac{1}{Q^2} \right)^{3/4} \equiv -\frac{3}{r_{\rm IR}^2 
\ell_{\rm pl}^2}.  
\nonu
\eea
One finds that
the fluctuation spectrum for $u, v, w$ fields around ${\cal N}=3$ fixed point
takes two positive values and one negative value:
\bea
M_{\overline{u} \overline{u}}^2(N^{0,1,0}_{\rm I})= + 4 \frac{1}{r_{\rm IR}^2}, 
\hskip0.5cm 
M_{\overline{v} \overline{v}}^2(N^{0,1,0}_{\rm I}) = +4 \frac{1}{r_{\rm IR}^2}, \hskip0.5cm 
M_{\overline{w} \overline{w}}^2(N^{0,1,0}_{\rm I}) = -2 \frac{1}{r_{\rm IR}^2}. 
\nonu
\eea
Via AdS/CFT correspondence, one finds that in $d=3$ conformal field theory
with ${\cal N}=3$ supersymmetry, the squashing ought to
be an irrelevant perturbation of conformal dimension $\De=4$ for $u, 
v$ fields and an relevant perturbation of conformal dimension $\De=2 $ or
$\De=1$ for $w$ field.
According to the result of \cite{df}, mass spectrum of the Lichnerowitz 
scalar field $\phi$ can be obtained from that of transverse spinor 
$\lambda_{\rm T}$ via $
m_{\phi}^2 = m_{\lambda_{\rm T}} ( m_{\lambda_{\rm T}}-4)$
in the normalization convention, $(\Box - 32 +m_\phi^2 ) \phi = 0$.
Moreover, from \cite{df1},  $m_{\lambda_{\rm T}} = {\cal D} +3$, 
where ${\cal D}$ is the Dirac operator of each spinor harmonics. 
Hence, canonically normalized mass of the Lichnerowitz scalar is given by
$\widetilde{m_{\phi}}^2 = -({\cal D} +3)({\cal D} -1)+32$. Unfortunately, 
at present, complete eigenvalues of the Dirac operator ${\cal D}$ is not known
for the transverse harmonics, even though there have been attempts recently 
to classsify the full mass spectra \cite{npqr}. 
We anticipate that the linear combination of masses of $u$ and $v$ 
fields should correspond to $\widetilde{m_{\phi}}^2$.

\vskip0.3cm
\tt
$\bullet$ Type II solution for the space $N_{\rm II}^{0,1,0}$
\rm
\vskip0.3cm

The Type II solution with ${\cal N} = (0, 1)$ supersymmetry exists, 
as pointed out in \cite{pp}, in the range of $c$:
\bea
\frac{2}{\sqrt{5}} \leq c \leq 1,  \qquad d=-\sqrt{1-c^2}.
\nonu
\eea
For $N^{0,1,0}_{\rm II}$, it leads to  
\bea
\al^2=\frac{400}{9}e^2, \quad \be=\pm \frac{20}{3} e, \quad
\ga^2= \frac{160}{9}e^2, \quad \de^2=\frac{160}{9}e^2 ,
\nonu
\eea
where we have put $c=1$ and $d=0$.
One again obtains
\bea
& & V_{uu}|= 189 \frac{2^{3/7}}{5^{10/7}} e^{9s_{\rm II}}, \qquad 
V_{vv}|=-6 \frac{2^{3/7}}{5^{10/7}} e^{9s_{\rm II}}, \qquad 
V_{ww}| =11 \frac{2^{3/7}}{5^{10/7}} e^{9s_{\rm II}}, \nonu \\
& &  V_{uv}|=-84 \frac{2^{3/7}}{5^{10/7}} e^{9s_{\rm II}}, \quad \quad
 V_{uw}|=0, \quad \quad V_{vw}|=0, 
\nonu
\eea
all evaluated at the extremum 
\bea
e^{12s_{\rm II}}=\frac{81}{2^{18/7} \times 5^{10/7}} \frac{1}{Q^2},
\qquad u_{\rm II}=\frac{2}{21} \ln \frac{2}{5}, \qquad 
v_{\rm II}= \frac{1}{3} \ln \frac{2}{5}, \qquad w_{\rm II}=0.
\nonu
\eea
From these, one calculates mass spectrum of the scalar fields 
straightforwardly:
\bea
M_{\overline{u} \overline{u}}^2(N^{0,1,0}_{\rm II}) & = 
& \left[ \frac{\pa^2 V}{\pa \overline{u}^2}\right]_{
\overline{s_{\rm II}},
\overline{u_{\rm II}}, 
\overline{v_{\rm II}},
\overline{w_{\rm II}}} \,\,\, = \qquad  
+ 9 \times \frac{2^{3/7}}{5^{10/7}} e^{9s_{\rm II}}, \nonu \\
M_{\overline{u} \overline{v}}^2(N^{0,1,0}_{\rm II}) & = 
& \left[ \frac{\pa^2 V}{\pa \overline{u} 
\pa \overline{v}} \right]_{
\overline{s_{\rm II}}, \overline{u_{\rm II}}, 
\overline{v_{\rm II}}, \overline{w_{\rm II}}} 
= \qquad -2 \sqrt{14} \times \frac{2^{3/7}}{5^{10/7}} 
e^{9s_{\rm II}},    \nonu \\
  M_{\overline{v}\overline{v}}^2(N^{0,1,0}_{\rm II}) & = & 
\left[ \frac{\pa^2 V}{\pa \overline{v}^2}\right]_{
\overline{s_{\rm II}},
\overline{u_{\rm II}}, \overline{v_{\rm II}},
\overline{w_{\rm II}}} \,\,\, =  \qquad - \frac{2^{3/7}}{5^{10/7}} 
e^{9s_{\rm II}},  \nonu \\
  M_{\overline{w} \overline{w}}^2(N^{0,1,0}_{\rm II}) & = & 
\left[ \frac{\pa^2 V}{\pa \overline{w}^2}\right]_{
\overline{s_{\rm II}}, \overline{u_{\rm II}}, \overline{v_{\rm II}},
\overline{w_{\rm II}}} \,\,\, = \qquad 
\frac{11}{2} \times \frac{2^{3/7}}{5^{10/7}} e^{9s_{\rm II}}.   
\nonu
\eea
Diagonalizing the mass matrix , one obtains th mass eigenvalues  
\bea
M^2 = -5  \times \frac{2^{3/7}}{5^{10/7}} e^{9s_{\rm II}} , \quad
+ 13  \times \frac{2^{3/7}}{5^{10/7}} e^{9s_{\rm II}}, \quad 
+\frac{11}{2} \times  \frac{2^{3/7}}{5^{10/7}} e^{9s_{\rm II}}.
\nonu
\eea
The eigenvector $\sqrt{2} \overline{u}+ \sqrt{7} \overline{v}=\sqrt{42}(u+v)$ 
(corresponding to the `tachyonic' eigenvalue) in fact represents squashing
of $N^{0,1,0}$ manifold, whose magnitude is parametrized by $\la^2$: 
\bea
\la^2= \frac{1}{4} e^{\frac{7}{3} (u+ v)}.
\nonumber 
\eea
For $N^{0,1,0}_{\rm I}$, $\lambda^2 = 1/2$, while, for $N^{0,1,0}_{\rm II}$,
$\lambda^2 = 1/10$. 

The cosmological constant of Type II solution  $\Lambda_{\rm II}$ is
\bea
\La_{\rm II}= \frac{V_{\rm II}}{2} = - \frac{729 \sqrt{10}}{2 \times
10^3} \left( \frac{1}{Q^2} \right)^{3/4} \equiv -\frac{3}{r_{\rm UV}^2 
\ell_{\rm pl}^2}.  
\nonu
\eea
One finds that
the fluctuation spectrum for $u, v, w$ fields around ${\cal N}=1$ fixed point
takes one negative value and two positive values:
\bea
M_{\widetilde{u} \widetilde{u}}^2(N^{0,1,0}_{\rm II}) =    
-\frac{20}{9} \frac{1}{r_{\rm UV}^2}, \qquad 
  M_{\widetilde{v}\widetilde{v}}^2(N^{0,1,0}_{\rm II})  =  
 \frac{52}{9} \frac{1}{r_{\rm UV}^2}, \qquad 
  M_{\overline{w} \overline{w}}^2(N^{0,1,0}_{\rm II})  =  
\frac{22}{9} \frac{1}{r_{\rm UV}^2} ,
\nonu
\eea
where $\widetilde{u} =\sqrt{\frac{14}{3}}(u+v)$ and $\widetilde{v}=
\frac{1}{\sqrt{3}}(7u-2v)$.

From the above mass spectrum, one finds that, in ${\cal N}=1$ superconformal 
field theory, the squashing deformation ought to be a relevant perturbation 
of conformal dimension $\De=5/3$ or $4/3$. Scaling dimensions of other 
deformations are $\Delta = 13/3$ for $\widetilde{v}$ field and 
$\Delta = 11/3$ for $w$ field, respectively.

\subsection{$SO(5)/SO(3)_{\rm max}, V_{5,2}({\rm R}), M^{1,1,1}, Q^{1,1,1},
N^{p,q,r}$ }
We will now study other homogeneous Einstein manifolds, for which
four-dimensional scalar potential is known on an appropriate subspace.
\vskip0.3cm
\tt
$\bullet N^{p,q\ne 0, r}_{\rm I, II}$
\rm
\vskip0.3cm
As recalled above, Page and Pope have shown that, for $q \ne 0$, by
geometric squashing, $N^{p,q,r}_{\rm I}$ manifold leads, much as in 
$N^{0,1,0}$'s, to $N^{p,q,r}_{\rm II}$. 
It is straightforward to show that the effective potential Eq.(4),
which is valid for all $x$ hence $(p,q \ne 0,r)$, have two critical points. 
The supersymmetry preserved by $N^{p,q,r}_{\rm I}$ and
$N^{p,q,r}_{\rm II}$ is ${\cal N}= (1,0)$ and ${\cal N}=(0,1)$, respectively.
On the other hand, depending on the value of $x$, relative magnitdue
of the cosmological constants, $V_{\rm I, II}$ , changes.  
Hence, in three-dimensional conformal field theory with $SU(3) \times 
U(1)$ global symmetry dual to $AdS_4 \times N^{p,q,r}_{\rm I, II}$'s, 
there will be two classes of renormalization group flows:
one flowing from a ${\cal N}=1$ UV fixed point to a nonsupersymmetric IR
fixed point, and another flowing from a nonsupersymmetric UV fixed point
to a ${\cal N}=1$ IR fixed point. 

\vskip0.3cm
\tt
$\bullet SO(5)/SO(3)_{\rm max}$
\rm
\vskip0.3cm 
The $SO(5)/SO(3)_{\rm max}$ seven-manifold is constructed from 
maximal embedding of $SO(3)$ in $SO(5)$. Let us  
consider squashing each coset direction as specified by the vielbein:
\bea
B^{\al}=(\frac{1}{\al} \Omega^a, \frac{1}{\be} \Omega^{\hat{a}}, 
\frac{1}{\ga} \Omega^0) ,
\nonu
\eea
where $\Omega^{a}( a=1, 2, 3), 
\Omega^{\hat{a}}(\hat{a}=1, 2, 3)$ and 
$\Omega^0$ are the left invariant one-forms
corresponding to the generators $X^a, X^{\hat{a}}, X^0$
of $SO(5)$, respectively. 
Parametrizing as  
\bea
\al=e^{s+(u+v)/2}, \;\; \be=e^{s+(u-v)/2}, \;\; \ga=e^{s-3u},
\nonu
\eea
the squashing deformation can be summarized compactly in terms of  
the following four-dimensional effective Lagrangian \cite{yasuda1}:
\bea
{\cal L}_4=\sqrt{-g_4} \left( R_4-\frac{63}{2} \left(
\pa_{\mu} s \right)^2-\frac{21}{2} \left(\pa_{\mu} u \right)^2-
\frac{3}{2} \left( \pa_{\mu} v \right)^2 -
V(s,u,v) \right) .
\nonu
\nonu
\eea
Here, the scalar potential is given by
\bea
V(s,u,v) & = & e^{9s} \left( -12 e^{u+v} -7 e^{u-v}-2 e^{-6u}+
e^{8u} +e^{-6u+2v} +e^{-6u-2v} +e^{u-3v} \right. \nonu \\
& & \left. -30 e^u -16 e^{-(5u+v)/2}
\right)+Q^2 e^{21s} .
\nonu
\eea
Again, $Q$ is the conserved `Page' charge. It is straightforward to check
that there exists only one critical point, $u=v=0$. In the subspace 
of the perturbations considered above, one finds that the strongly interacting 
$d=3$ quantum field theory dual to $AdS_4 \times SO(5)/SO(3)_{\rm max}$ has 
only one isolated (super)conformal fixed point, hence, no nontrivial 
renormalization group flow.  

\vskip0.3cm 
$\bullet V_{5,2}({\rm R})$
\vskip0.3cm
The $V_{5, 2} ({\rm R})$ is a real Stiefel manifold, constructed by
embedding $SO(3)$ in $SO(5)$ in such a way that $\bf 5$ of $SO(5)$ 
branches into $\bf 3$ of $SO(3)$ plus two singlets. The isometry of
$V_{5, 2}(R)$ is $SO(5) \times SO(2)$. Consider deformations rescaling 
each coset direction as 
\bea
B^{\al}=(\frac{1}{\al} \Omega^m, \frac{1}{\be} \Omega^{\hat{m}}, 
\frac{1}{\ga} \Omega^0),
\nonu
\eea
where $\Omega^{m}( m=1, 2, 3), 
\Omega^{\hat{m}}(\hat{m}=1, 2, 3)$ and 
$\Omega^0$ are the left invariant one-forms
corresponding to the generators $X^m=T^{4m}, X^{\hat{m}}=T^{5m}, X^0=T^{45}$
where $T^{ij}(i,j=1,2,3,4,5)$
are generators of $SO(5)$. 
In terms of parametrization  
\bea
\al=e^{s+(u+v)/2}, \;\; \be=e^{s+(u-v)/2}, \;\; \ga=e^{s-3u},
\nonu
\eea
the four-dimensional effective Lagrangian \cite{yasuda1} can be 
written as
\bea
{\cal L}=\sqrt{-g_4} \left( R_4-\frac{63}{2} \left(
\pa_{\mu} s \right)^2-\frac{21}{2} \left(\pa_{\mu} u \right)^2-
\frac{3}{2} \left( \pa_{\mu} v \right)^2 -
V(s,u,v) \right) ,
\nonu
\eea
where 
\bea
V(s,u,v) & = & e^{9s} \left( -6 e^{u+v} -6 e^{u-v}-2 e^{-6u}+
e^{-6u+2v} +e^{-6u-2v} + e^{8u}
\right)  \nonu \\
& & +Q^2 e^{21s}
\nonu
\eea
and $Q$ is the Page charge. The potential exhibits only one critical point at 
$u=\frac{1}{7} \ln \frac{3}{2}, v=0$. As in  $SO(5)/SO(3)_{\rm max}$ case, 
there is no renormalization group flow in the corresponding three-dimensional 
conformal field theory.
\vskip0.3cm
$\bullet M^{1,1,1}$
\vskip0.3cm
The manifold $M^{1,1,1}$ is a homogeneous space 
$[SU(3) \times SU(2) \times U(1)]/[SU(2) \times U(1) \times U(1)]$. 
For the vielbein $B^{\al}$, we rescale each coset direction as 
\bea
B^{\al}=(\frac{1}{b} \Omega^m, \frac{1}{c}( \sqrt{3} \Omega^8+ \Omega^3 +
2 \Omega^{3'}), \frac{1}{a} \Omega^{A}) 
\nonu
\eea
where $\Omega^{m}( m=1, 2), \Omega^3, \Omega^8, \Omega^{A} (A=4, 5, 6, 7), 
$ and $\Omega^{3'}$ are the left invariant one-forms
corresponding to the coset generators $\si^m$ and 
$\si^3$ of $SU(2)$, $\la^8$ and $\la^A$ of $SU(3)$,
and the generator of $U(1)$ factor respectively.
In terms of parametrization  
\bea
a=e^{s+(u-v)/2}, \;\; b=e^{s+u/2+v}, \;\; c=e^{s-3u},
\nonu
\eea
the four-dimensional effective Lagrangian \cite{yasuda} can be written as
\bea
{\cal L}=\frac{1}{2} \sqrt{-g_4} \left( -R_4-\frac{63}{2} \left(
\pa_{\mu} s \right)^2-\frac{21}{2} \left(\pa_{\mu} u \right)^2-
3 \left( \pa_{\mu} v \right)^2  -
V(s,u,v) \right).
\nonu
\eea
The scalar potential is given by
\bea
V(s,u,v) & = & e^{9s} \left( -3 e^{u-v}- e^{u+2v} + \frac{9}{8} e^{8u-2v}+
\frac{1}{4} e^{8u+4v} \right)+Q^2 e^{21s},
\nonu
\eea
where again $Q$ is the conserved Page charge.
The stationary conditions 
$\frac{\pa V}{\pa s}= \frac{\pa V}{\pa u}= \frac{\pa V}{\pa v}=0$ lead to a
set of cubic equations for $t \equiv 3 e^{-3 v}$ :
\bea
t^3-3 t^2+4t-4=(t-2)(t^2-t+2)=0 .
\nonu
\eea 
One finds that there is no extra critical point except $t=2$.
Hence, one concludes that there ought to be no nontrivial renormalization 
group flow in the dual, $d=3$ (super)conformal field theory.
\vskip0.3cm
$\bullet Q^{1,1,1}$
\vskip0.3cm 
The manifold $Q^{1,1,1}$ is a homogeneous space
$[SU(2) \times SU(2) \times SU(2)]/[U(1) \times U(1) \times U(1)]$. 
For the vielbein $B^{\al}$, we rescale each coset direction as 
\bea
B^{\al}=(\frac{1}{a} \Omega^i, \frac{1}{d}(  \Omega^0+ \Omega^{0'} +
 \Omega^{0''}), \frac{1}{b} \Omega^{i'}, \frac{1}{c} \Omega^{i''}) 
\nonu
\eea
where $\Omega^{i}( i=1, 2), \Omega^0, \Omega^{i'} (i'=4, 5), 
\Omega^{0'}, \Omega^{i''}(i''=6, 7)$ and $\Omega^{0''}$ 
are the left invariant one-forms
corresponding to the coset generators $\si^i$ and $\si^3$ of $SU(2),
SU(2)$ and$SU(2)$ respectively. 
By parametrizing  
\bea
a=e^{s+(u+v+w)/2}, \;\; b=e^{s+(u+v-w)/2}, \;\; c=e^{s+u/2-v}, \;\;
d=e^{s-3u},
\nonu
\eea
the four dimensional effective Lagrangian \cite{yasuda} can be 
written as
\bea
{\cal L}=\frac{1}{2} \sqrt{-g_4} \left( -R_4-\frac{63}{2} \left(
\pa_{\mu} s \right)^2-\frac{21}{2} \left(\pa_{\mu} u \right)^2-
3 \left( \pa_{\mu} v \right)^2 -\left( \pa_{\mu} w \right)^2 -
V(s,u,v,w) \right) ,
\nonu
\eea
where 
\bea
V(s,u,v,w) & = & e^{9s} \left( -e^{u+v+w}-e^{u+v-w}-e^{u-2v} +
\frac{1}{4} e^{8u+2v+2w}+\frac{1}{4} e^{8u+2v-2w}
+\frac{1}{4} e^{8u-4v} \right) \nonu \\
& & +Q^2 e^{21s} .
\nonu
\eea
The stationary conditions 
 $\frac{\pa V}{\pa u}= \frac{\pa V}{\pa v}=
\frac{\pa V}{\pa w}=0$ yield a set of simple algebraic equations
\bea
\al(1-\al)=\be(1-\be)=\ga(1-\ga)=\al^2+\be^2+\ga^2 ,
\nonu
\eea
where
$\al \equiv \frac{a^2}{2d^2}, \be \equiv \frac{b^2}{2d^2}, \ga \equiv 
\frac{c^2}{2d^2}$. The only solution is 
\bea
2\al=1-\sqrt{1-t}=2\be=2\ga, \;\;\;\;\;
3-3\sqrt{1-t}-2t=0.
\nonu
\eea
It is easy to see that there exists only one extremum value when $t=3/4$. 
One again concludes that there is no nontrivial renormalization group flow 
in the dual, $d=3$ (super)conformal field theory. 

\bea
\begin{array}{|c|c|c|c|c|}
\hline 
X_7 & \mbox{Isometry}  & \mbox{Holonomy}  & \mbox{Supersymmetry} & 
\mbox{RG flow} \nonu \\
\hline
\mbox{Round} \;\; \S^7 & SO(8) &1 &(8,8) & \mbox{Yes}^\sharp \nonu \\
\hline
\mbox{Squashed} \;\; \S^7 & SO(5) \times SU(2)& G_2&1 & \mbox{Yes}^\sharp 
\nonu \\
\hline
SO(5)/SO(3)_{max}&SO(5) & G_2 & (1,0) & \mbox{No} \nonu \\
\hline
V_{5,2}({\rm R}) & SO(5) \times U(1) & SU(3) & (2,0) & \mbox{No} \nonu \\
\hline
M^{1, 1, 1} & SU(3) \times SU(2) \times U(1) & SU(3) & (2,0) & \mbox{No}\nonu\\
\hline
M^{p, q, r} & SU(3) \times SU(2) \times U(1) & SO(7)  & 0 & 
 \nonu \\
\hline
Q^{1, 1, 1}& SU(2)^3 \times U(1) & SU(3) & (2,0) & \mbox{No} \nonu \\
\hline
Q^{p, q, r} &  SU(2)^3 \times U(1) & SO(7) &  0 &  \nonu \\
\hline
N^{0, 1, 0}_{\rm I} & SU(3) \times SU(2) & SU(2) & (3,0) & \mbox{Yes}^\star \nonu\\
\hline
N^{0, 1, 0}_{\rm II} & SU(3) \times SU(2) & G_2 & (0,1) & \mbox{Yes}^\star 
\nonu \\ \hline
N^{p, q \ne 0, r}_{\rm I} & SU(3) \times U(1) & G_2 & (1,0) & 
\mbox{Yes}^\dagger \nonu \\ 
\hline 
N^{p,q \ne 0,r}_{\rm II} & SU(3) \times U(1) & G_2 & (0,1) & 
\mbox{Yes}^\dagger \\ 
\hline
\end{array}
\eea

Table 1:
\sl Classification of Einstein spaces $X_7$ and Renormalization Group Flows.
The flow found in \cite{ar} is marked $\sharp$, while the ones found in 
Section 2.1 and 2.2 are marked $\star$ and $\dagger$.  \rm
\vskip0.4cm

We have summarized our result of this Section in Table 1. In the subspace
of squashing deformations considered, among various known $X_7$'s, we have 
found that only $N^{p,q \ne 0, r}_{\rm I, II}$'s turn out to be dual to
nontrivial renormalization group flows of strongly coupled three-dimensional
field theories, connecting two (super)conformal fixed points. One novelty of 
these conformal field theories is that, even though the superconformal 
symmetry is changed, the global symmetry is not changed at all, in contrast
to the deformation interpolating between round- and squashed-$\S^7$ 
\cite{ar}.

Although we have not considered nonsupersymmetric seven-manifolds 
$M^{p, q, r}$ or $Q^{p, q, r}$ in this paper, it is known that $M^{p, q, r}$ 
remains stable for the  specific region of $98/243 \leq p^2/q^2 
\leq 6358/4563$ and $Q^{p, q, r}$ solutions are stable in a certain region
containing the point $p=q=r=1$. We therefore anticipate that, if there 
exists more than one critical points of the corresponding scalar potential,
they will provide gravity dual to strongly interacting, stable, 
nonsupersymmetric field theories in three dimensions, 
whose renormalization group flows interpolate interacting conformal fixed 
points \cite{berkoozrey}. 

\section{ 3d CFTs from Englert Compactifications}
So far, we have deduced existence of three-dimensional conformal field 
theories out of Freund-Rubin compactification of M-theory. Let us now 
consider more general compactifications by relaxing the restriction that 
the eleven-dimensional metric is a product space and that there is no 
magnetic four-form field strength flux threaded on $X_7$. 
The first solution of this sort has been found by Englert
\cite{englert}. In contrast to the Freund-Rubin compactifications, the
symmetry of the vacuum is no longer given by the isometry group of
$X_7$ but rather by the group which leaves invariant both the metric
$g_{ab}$ and four-form magnetic field strength $G_{abcd}$. In the 
Englert compactification \cite{englert},  nonvanishing  $G_{abcd}$ on the 
round $\S^7$ breaks $SO(8)$ down to $SO(7)$. 
Precise interpretation of Englert compactification in terms of microscopic
configuration of coincident M2-branes is still lacking. Nevertheless, 
AdS/CFT correspondence implies that there ought to be a three-dimensional 
(super)conformal field theory for each Englert-type compactification as well.   

In Kaluza-Klein supergravity, it is well-known that the four-dimensional 
${\cal N}=8$ gauged supergravity can be embedded consistently into the 
eleven-dimensional supergravity. As shown in \cite{cj}, the 70 scalars of 
${\cal N}=8$ supergravity live on the coset space $E_7/SU(8)$ and are 
described by an element ${\cal V}(x)$ of the fundamental 56-dimensional 
representation (56-bein) of $E_7$ :
\bea
{\cal V}(x)=
\left[
\begin{array}{cc}
u_{ij}^{IJ}(x) & v_{ijKL}(x)  \\
v^{klIJ}(x) & u^{kl}_{KL}(x) 
\end{array}
\right] ,
\nonu
\eea
where $SU(8)$ index pairs $[ij], \cdots$ and $SO(8)$ index pairs $[IJ], \cdots$
are antisymmetrized and therefore $u$ and $v$ are $28 \times 28$ matrices.
Complex conjugation is done by raising or lowering indices, for example,
$(u_{ij}^{IJ})^{\star}= u^{ij}_{IJ}$.
Under local $SU(8)$ and local $SO(8)$, the matrix ${\cal V}(x)$ transforms
as ${\cal V}(x) \rightarrow U(x) {\cal V}(x) O^{-1}(x)$, where 
$U(x) \subset SU(8)$ and $O(x) \subset SO(8)$ and matrices $U(x)$,
$O(x)$ are elements of the 56-dimensional representation.
By appropriate gauge fixing of the local $SU(8)$ symmetry, 
the 56-bein ${\cal V}(x)$ can be brought into the following form:
\bea
{\cal V}(x)=
\mbox{exp} \left[\begin{array}{cc}
0 & -\frac{\sqrt{2}}{4} \phi_{ijkl}  \\
-\frac{\sqrt{2}}{4} \phi^{mnpq} & 0 
\end{array} \right] ,
\nonu
\eea  
where $\phi^{ijkl}$ is a complex self-dual tensor describing the 35 
scalars $\bf 35_{v}$(the real part of $\phi^{ijkl}$)
and 35 pseuoscalar fields $\bf 35_{c}$(the imaginary part of $\phi^{ijkl}$)
of the ${\cal N}=8$ supergravity. Note that, after gauge fixing, there is no
distinction between $SO(8)$ and $SU(8)$ indices. 

The scalar potential of the gauged ${\cal N}=8$ supergravity is known
to possess four critical points with at least $G_2$ invariance \cite{warner}.
The maximally supersymmetric vacuum with $SO(8)$ symmetry, $\S^7$, 
is where expectation value of both scalar and pseudoscalar fields vanish.
Let us deonte self-dual and anti-self-dual tensors of $SO(8)$ tensor as
$C_{+}^{IJKL}$ and $C_{-}^{IJKL}$, respectively, satisfying
\bea
C_{\pm}^{IJMN} C_{\pm}^{MNKL} = 12 \de_{KL}^{IJ}  \pm 4 C_{\pm}^{IJKL}.
\nonu
\eea
Turning on the scalar fields proportional to $C_{+}^{IJKL}$ yields an 
$SO(7)^{+}$ invariant vacuum. Likewise, turning on pseudoscalar fields 
proportional to $C_{-}^{IJKL}$ yields $SO(7)^{-}$ invariant vacuum. 
Both $SO(7)^{\pm}$ vacua are nonsupersymmetric. However, simulatneously 
turning on both scalars and pseudoscalar fields proportional to 
$C_{+}^{IJKL}$ and $C_{-}^{IJKL}$, respectively, one obtains $G_2$-invariant 
vacuum with ${\cal N}=1$ supersymmetry \footnote{The $G_2$ is the common
subgroup of $SO(7)^+$ and $SO(7)^-$.}.  
The most general vacuum expectation value of 56-bein retaining 
$G_2$-invariance can be parametrized as
\bea
\langle \phi_{IJKL} \rangle 
=\frac{\la}{2 \sqrt{2}} \left( \cos \al \; C_{+}^{IJKL}
+i \sin \al \; C_{-}^{IJKL} \right).
\label{vev}
\nonu
\eea
In this case, the elements of 56-bein ${\cal V}(x)$ can be written as
:
\bea
u^{IJ}_{KL}(\la) & = & 2 p^3 \; \de_{KL}^{IJ} + 
\frac{1}{2}(1+\cos 2\al) p q^2 \;
C_{+}^{IJKL}+ \frac{1}{2}(1-\cos 2 \al) p q^2 \; C_{-}^{IJKL} \nonu \\
& & - i p q^2 \sin 2 \al \; D_{-}^{IJKL}, \nonu \\
v^{IJKL}(\la) & = & \frac{1}{2} ( 3 e^{i \al} + e^{-3 i \al}) q^3 \;
\de^{IJ}_{KL}+
p^2 q \cos \al \; C_{+}^{IJKL}- i p^2 q \sin \al \;  C_{-}^{IJKL} \nonu \\
& & +
\frac{1}{2}(e^{i \al}-e^{-3 i \al}) q^3 \; D_{+}^{IJKL}
,
\label{uv}
\eea 
where
$D^{\pm}_{IJKL} \equiv \frac{1}{2} \left( C_{+}^{IJMN} 
C_{-}^{MNKL} \pm C_{-}^{IJMN} C_{+}^{MNKL} \right), p \equiv 
\cosh(\la/2\sqrt{2})$
and $q \equiv \sinh(\la/2\sqrt{2})$.

\begin{figure}[htb]
\label{fig2}
\vspace{0.5cm}
\epsfysize=8cm
\epsfxsize=8cm
\centerline{
\epsffile{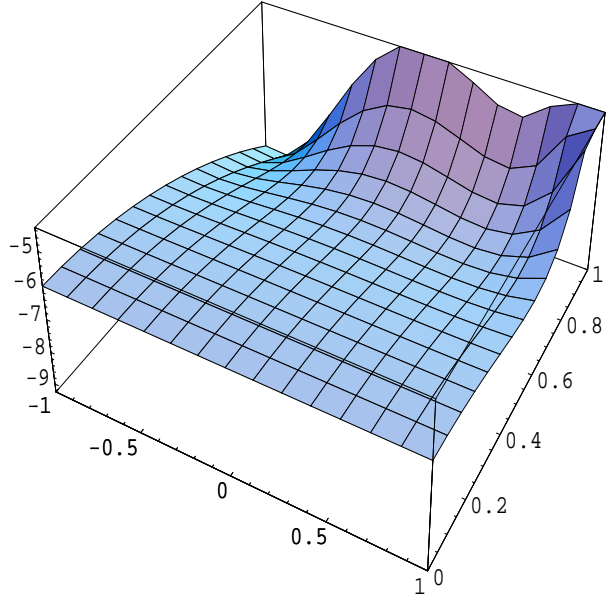} }
\vspace{0.5cm}
\caption{\sl Scalar potential $V(\al, \la)$. The left axis corresponds to
$v=\cos \al$ and right one does $\la$. The extremum value $V=-7.19$
for $G_2$ occurs
around $v=0.56$ and $\la=0.73$ while the extremum value $V=-6$ for $SO(8)$
appears around
$\la=0$. We take $g^2$ as 1 for simplicity. }
\end{figure}
\vspace{0.5cm}

In the above parametrization, the scalar potential is given by \cite{dnw}
\bea
V(\al, \la)=
2 g^2 \left( (7 v^4-7 v^2 +3) c^3 s^4 +(4v^2-7) v^5 s^7 + c^5 s^2 +
7 v^3 c^2 s^5 -3 c^3 \right)
\label{potential}
\eea
where $c \equiv \cosh (\la/\sqrt{2}), 
s \equiv \sinh(\la/\sqrt{2})$ and $v \equiv \cos \al$.
The potential is plotted in Figure 2 and the four critical points are
summarized in Table 2.  

\bea
\begin{array}{|c|c|c|c|c|}
\hline 
$\mbox{Gauge symmetry}$ & $\mbox{Supersymmetry}$ 
& \cos \al  & c^2 & V \nonu \\
\hline
SO(8) & {\cal N}=8 &    & 1 & - 6g^2 \nonu \\
\hline
SO(7)^{-} & {\cal N}=0  &   0  & 
\frac{5}{4} & -\frac{25 \sqrt{5}}{8} g^2=-6.99 g^2 \nonu \\
\hline
SO(7)^{+} & {\cal N}=0   & 1  & 
\frac{1}{2} (\frac{3}{\sqrt{5}}+1) &  -2 \cdot 5^{3/4} g^2 = -6.69 g^2 
\nonu \\
\hline
G_2 & {\cal N}=1 &
\frac{1}{2} \sqrt{ (3-\sqrt{3})}=0.56  &
\frac{1}{5}(3+2\sqrt{3}) & -\frac{216 \sqrt{2}}{25 \sqrt{5}} 3^{1/4} g^2 
= - 7.19 g^2 \nonu \\
\hline
\end{array}
\eea

Table 2. \sl Summary of four critical points: symmetry group, supersymmetry, 
vacuum expectation values of scalar and pseudoscalar fields, and cosmological
constants. 
\rm 


In this section, we will be identifying a renormalization group flow 
associated with global symmetry breaking $SO(8) \rightarrow G_2$
in a three-dimensional strongly coupled field theory \footnote{As the 
$SO(7)^\pm$ vacua are unstable \cite{dn1}, 
we will not consider in the foregoing discussions.}. We will show that 
the perturbation operator is relevant at the $SO(8)$ invariant UV fixed 
point but becomes irrelevant at the $G_2$ invariant IR fixed point.
To identify conformal field theory operator corresponding to the
perturbation while preserving $G_2$ symmetry, we will consider harmonic 
fluctuations of spacetime metric and $\la$ scalar field around 
$AdS_4 \times \S^7$. From the scalar potential Eq. (\ref{potential}),
one finds that the cosmological constant $\Lambda$ is given by
\bea
\La_{SO(8)}= - 6 g^2 \equiv -\frac{3}{r_{\rm UV}^2 \ell_{\rm pl}^2}  ,
\nonu
\eea
where $r_{\rm UV}$ is the radius of $AdS_4$ and $\ell_{\rm pl}$ 
is the eleven-dimensional Planck scale.
Conformal dimension of the perturbation operator representing this 
deformation
is calculated by fluctuation spectrum of the scalar fields. 
The kinetic term \cite{dn2} for $\lambda$ field can be obtained from
$-|A_{\al}^{ijkl}|^2/96$, where
\bea
A_{\al}^{ijkl} = -2 \sqrt{2} ( u^{ij}_{IJ} \partial_{\al} v^{klIJ} -
v^{ijIJ} \partial_{\al} u^{kl}_{IJ} ).
\nonu
\eea
Then, from the explicit forms of $u^{IJ}_{KL}$ and $v^{IJKL}$ in 
Eq. (\ref{uv}), one obtains $|A_{\al}^{ijkl}|^2=84 
(\partial_{\al} \lambda)^2$, where we have kept to quadratic order in 
the fluctuation of $\lambda$. The resulting kinetic term is 
$-\frac{7}{8}(\partial_{\al} \lambda)^2$. After rescaling the $\lambda$ 
field as $\overline{\lambda}= \sqrt{\frac{7}{4}} \lambda$, one finds that
the mass spectrum of the $\overline{\la}$ field around $SO(8)$ fixed point
is given by:
\bea
 \pa_{\overline \la}^2 
V(SO(8))\Big|_{\overline{\la}=0}  =  -2 g^2 \ell_{\rm pl}^2 
= -2 \frac{1}{r_{\rm UV}^2 }. 
\label{m^2}
\eea
Via AdS/CFT correspondence, one finds that in the corresponding ${\cal N}=8$
superconformal field theory, the $G_2$ symmetric deformation ought to
be a relevant perturbation of conformal dimension $\De=1$ or $\De=2$.
Recall that, on $\S^7$, mass spectrum of the representation corresponding
to $SO(8)$ Dynkin label $\bf (n, 0, 2, 0)$ is given by 
\bea
\widetilde{M^2}= \left( (n+1)^2-9 \right) m^2
\nonu
\eea
where $m^2$ is mass-squared parameter of a given $AdS_4$ spacetime and
a scalar field $S$ is defined by $(\De_{\mbox{AdS}}+\widetilde{M^2})S=0$.
This follows from the known mass formula \cite{bcers} $M^2=((n+1)^2-1) m^2$
for $O^{-(1)}$ and the fact that $M^2$ is traditionally defined according to
$(\De_{\mbox{AdS}}-8 m^2 +M^2)=0$. For $\bf 35_c$ corresponding to $n=0$,
$\widetilde{M^2}_{\bf 35_{c}}=-8 m^2$ and this ought to equal to Eq.(\ref{m^2}).
Recalling that $r_{\rm UV}^2= r_{\S^7}^2/4=1/4m^2$,
\bea
 \pa_{\overline \la}^2 V(SO(8)) \Big|_{\overline{\la}=0} =
 -2 \frac{1}{r_{\rm UV}^2 }= \widetilde{M^2}_{\bf 35_{c}}
\nonu 
\eea
On the other hand, the mass spectrum of the representation corresponding
to $SO(8)$ Dynkin label $\bf (n+2, 0, 0, 0)$ is given by $
\widetilde{M^2}= \left( (n-1)^2-9 \right) m^2$.
This follows from the known mass formula \cite{bcers} $M^2=((n-1)^2-1) m^2$
for $O^{+(1)}$.
 For $\bf 35_v$ corrsponding to $n=0$,
$\widetilde{M^2}_{\bf 35_{v}}=-8 m^2$ and this ought to equal to
Eq. (\ref{m^2}).

Let us next consider the conformal fixed point corresponding to the 
$G_2$ symmetry. Again, from the scalar potential Eq.(\ref{potential}),
one finds that cosmological constant $\Lambda$ is given by
\bea
\La_{G_2}= -  \frac{216 \sqrt{2}}{25 \sqrt{5}}
3^{1/4} g^2  \equiv -\frac{3}{r_{\rm IR}^2 \ell_{\rm pl}^2}.  
\nonu
\eea
The mass spectrum for the $\overline{\la}$ field around $G_2$ fixed point
takes a positive value:
\bea
 \pa_{\overline \la}^2 V(G_2) \Big|_{c^2=\frac{1}{5}(3+2\sqrt{3})}  =  
15.446 g^2 \ell_{\rm pl}^2 = 6.443 \frac{1}{r_{\rm IR}^2 }. 
\nonu
\eea
One finds that in the corresponding three-dimensional conformal field theory
with ${\cal N}=1$ supersymmetry, the $G_2$ symmetric deformation ought to
be an irrelevant perturbation of conformal dimension $\De=4.448...$.
We thus conclude that the perturbation operator dual to the $\lambda$
field induces nontrivial renormalization group flow from ${\cal N}=8$
superconformal UV fixed point with $SO(8)$ symmetry to ${\cal N}=1$ 
superconformal IR fixed point with $G_2$ symmetry.


It is known that, in ${\cal N}=8$ supergravity, there also exists 
a ${\cal N}=2$ supersymmetric, $SU(3) \times U(1)$ invariant vacuum \cite{nw}. 
To reach this critical point, one has to turn on expectation values of both 
scalar and pseudoscalar fields as
\bea
\langle \phi_{IJKL} \rangle =\frac{1}{2 \sqrt{2}} \left( \la \; X^{+}_{IJKL}
+i \la' \; X^{-}_{IJKL} \right),
\nonu
\eea 
where
\bea
  X^{+}_{ijkl} &=& +[ (\de^{1234}_{ijkl}+\de^{5678}_{ijkl})+
 (\de^{1256}_{ijkl}+ \de^{3478}_{ijkl})+(\de^{1278}_{ijkl}
 +\de^{3456}_{ijkl})]
\nonu \\
       X^{-}_{ijkl} &=& -[(\de^{1357}_{ijkl}
-\de^{2468}_{ijkl})+(\de^{1268}_{ijkl}
     -\de^{2457}_{ijkl})+(\de^{1458}_{ijkl} -\de^{2367}_{ijkl})-
   (\de^{1467}_{ijkl}-\de^{2358}_{ijkl})]
\nonu
\eea
The scalar potential is
\bea
V & = & \frac{1}{2} g^2 \left( s'^4 \left( (x^2+3) c^3 + 4x^2 v^3 s^3 -
3v(x^2-1) s^3 + 12 x v^2 c s^2 - 6(x-1) c s^2 + 6(x+1) c^2 s v\right) 
\right. \nonu \\
& & \left. +
2s'^2 \left( 2(c^3+v^3 s^3) + 3 (x+1) v s^3 +6 x v^2 c s^2 -3(x-1)
c s^2 -6c\right) -12 c  \right)
\nonu
\eea
where
\bea
& & c \equiv \cosh(\la/\sqrt{2}), \;\; s \equiv \sinh(\la/\sqrt{2}), \;\;
c' \equiv \cosh(\la'/\sqrt{2}), \;\; s' \equiv \sinh(\la'/\sqrt{2}) \nonu \\
& & v \equiv \cos\al, \;\; x = \cos 2{\cal \phi}
.
\nonu
\eea
At the critical point, $s=1/\sqrt{3}, s'=1/\sqrt{2}$ and $\al=0$, 
one finds that the cosmological constant is given by 
$
V= -\frac{9 \sqrt{3}}{2} g^2 .
\nonu
$ 
Hence, one expects that there ought to exist a renormalization group 
flow between ${\cal N}=8 $ $SO(8)$ fixed point and  ${\cal N}=2$ 
$SU(3) \times U(1)$ fixed point. 

\section{6d CFTs from Inhomogeneous Compactification}
We finally study the case of near-horizon geometry of coincident M5-branes,
as described by $AdS_7$ gauged supergravity. Recently, in \cite{lp},  
geometrical interpretation for vacua in seven-dimensional ${\cal N}=1$ 
gauged supergravity has been given in terms of Englert type compactification 
of eleven-dimensional supergravity, having nonvanishing electric or tilted
magnetic four-form fluxes and {\sl inhomogeneous} metric deformation, by 
a set of consistent {\sl nonlinear} ansatz. 
The inhomogeneous deformations of the $\S^4$ is parametrized by a scalar ,
field $\phi$, while the $SU(2)$ gauge fields, which are the surviving 
subgroup of the $SO(5)$ Yang-Mills fields of the maximal gauged supergravity
after the deformation, are associated with the right translations under the 
$SU(2)$. The full two parameter potential of the resulting seven-dimensional 
gauged supergravity turns out equal to \cite{mtv} (See also \cite{st}):
\bea
V(\phi)=16 \; h^2 \; e^{-\frac{8}{\sqrt{5}} \phi} \,\, - \,\, 
8 \sqrt{2} \; h \; g \; e^{-\frac{3}{
\sqrt{5}} \phi} \,\, - \,\, g^2 \; e^{\frac{2}{\sqrt{5}} \phi} ,
\label{ads7pot}
\eea
where $h$ and $g$ are arbitrary real constants and $ e^{-\frac{1}{\sqrt{5}} 
\phi}$ represents the $SU(2)$ gauge coupling constant.
As shown in Figure 3, provided $h / g > 0$, the scalar potential has two 
extrema.

\begin{figure}[htb]
\label{fig3}
\vspace{0.5cm}
\epsfysize=6cm
\epsfxsize=8cm
\centerline{
\epsffile{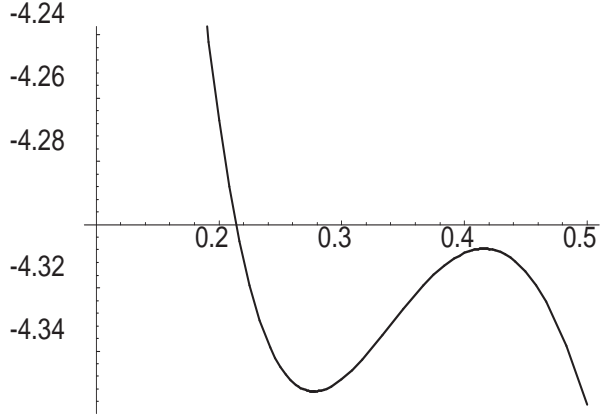} }
\vspace{0.5cm}
\caption{\sl Schematic shape of the scalar potential of $d=7$, ${\cal N}=1$ 
gauged supergravity. }
\end{figure}

A local maximum of the potential Eq.(\ref{ads7pot}) is located at 
\bea
\phi_{\rm I}= \frac{1}{\sqrt{5}} \ln \frac{8 \sqrt{2} \; h}{g} ,
\nonu
\eea
whose curvature equals to
\bea
\left[ \frac{\pa^2 V}{\pa \phi^2 } \right]_{\phi=\phi_{\rm I}}  
= -2 \times 2^{1/5} g^2 \left( \frac{\sqrt{2} \; h}{ g} \right)^{2/5} \equiv
-8 \frac{1}{r_{\rm UV}^2}.
\label{mass}
\eea
One then infers the radius of $AdS_7$ \footnote{
The radius of $AdS_7$ space is defined by $R_{\al \be \ga \de}
=- \frac{1}{r_{\rm UV}^2} (g_{\al \ga} g_{\be \de} -g_{\al \de} g_{\be \ga})$.}
as
\bea
r_{\rm UV}^2= \frac{4}{(2 h/g)^{2/5} g^2 } = -15/V_{\rm I}
\quad {\rm where} \quad
V_{\rm I}= -\frac{15}{2 \times 2^{4/5}} 
g^2 \left( \frac{\sqrt{2} h}{g} \right)^{2/5}.
\nonu
\eea
According to the analysis of \cite{bf}, if scalar field fluctuation 
about the maximum of the  potential $V$ satisfy  
$(\Box + \frac{\al}{r_{\rm UV}^2})\phi =0$, then perturbative stability is 
guaranteed provided $\al \leq (d-1)^2/4$. In the present case, the stability 
bound yields $\al_{s}=9$ when $d=7$. Eq.(\ref{mass}) is less than the bound, 
hence, is stable. In fact, the vacuum is ${\cal N}=1$ supersymmetric.
Conformal dimension of the perturbation operator representing the squashing
is calculated by fluctuation spectrum of the scalar fields. 
From Eq.(\ref{mass}), via AdS/CFT correspondence, one finds that, in 
six-dimensional ${\cal N}=(1,0)$ superconformal field theory, the perturbation 
operator representing the scalar field deformation ought to be a relevant 
perturbation of conformal dimension $\De=2$ or $\De=4$.

A local minimum of the scalar potential Eq.(\ref{ads7pot}) is located at
\bea
\phi_{\rm II}= \frac{1}{\sqrt{5}} \ln \frac{4 \sqrt{2} \; h}{g} ,
\nonu
\eea
around which
\bea
\left[ \frac{\pa^2 V}{\pa \phi^2 } \right]_{\phi=\phi_{\rm II}}  
= 2 \times 2^{4/5} g^2 \left( \frac{\sqrt{2} \; h}{ g} \right)^{2/5} \equiv
+12 \frac{1}{r_{\rm IR}^2}.
\nonu
\eea
One finds that the radius of $AdS_7$ is changed to :
\bea
r_{\rm IR}^2 = -15/V_{\rm II}  
\qquad {\rm where} \qquad
V_{\rm II}= - 
\frac{5}{2^{1/5}} g^2 \left( \frac{\sqrt{2} h} {g}  \right)^{2/5}.
\nonu
\eea
Dual to the critical point is a six-dimensional conformal field theory
with no supersymmetry, for which the perturbation associated with the 
squashing deformation is an irrelevant perturbation of conformal dimension 
$\De=3 + \sqrt{21}$.

One thus finds that, in the subspace representing inhomogeneous deformation 
of $\S^4$, there ought to be  a six-dimensional quantum theory 
which interpolates between ${\cal N}=(1,0)$ superconformal fixed point 
in the UV and nonsupersymmetric fixed point in the IR. Note that 
the coformal field theory around the IR fixed point is a stable one, 
despite absence of any supersymmetry, which follows from the general
argument of \cite{berkoozrey}. Existence of a stable, nonsupersymmetric
$d=6$ conformal field theory would be of considerable interest, 
as the theory would contain {\sl stable, noncritical bosonic strings} as 
part of the spectra. 

\vspace{1cm}
\centerline{\bf Acknowledgments} 
We are grateful to H. Nicolai and O. Yasuda for helpful correspondences on 
their works.

\end{document}